# iSSEE: IMS Sensors Search Engine Enabler for Sensors Mash-ups Convergent Application

Abdelkader Outtagarts and Olivier Martinot

Alcatel-Lucent Bell Labs
Centre de Villarceaux - Route de Villejust
91620 Nozay, France

**Abstract**
Integrating the sensing capabilities in Internet Protocol network will open the opportunities to build a wide range of novel multimedia applications. The problem when using sensors (e.g. temperature sensor, camera, audio, humidity, etc …) connected to the network is to know dynamically at any time if they are always connected or not, what type of data they can transmit and where they are geographically located. This paper describes an application enabler: IMS Sensor Search Engine Enabler (iSSEE), which allows IMS applications using sensors and IMS based devices, to get information about the sensor availability, its location and the type of the sensor. Using data collected by sensors and from the web, mash-ups convergent applications use cases are proposed by combining the contents from heterogeneous data.

## 1. Introduction

Sensor nodes (image, video, temperature, audio, etc…) are used in a wide range of applications to collect information. These information can be restituted to an end-user on his terminal or used to remotely control devices. We can also use these data collections to build multimedia mash-up applications which allow enriching services provided to end users. Before presenting the related works on sensors based applications, lets know more about IP Multimedia Subsystem (IMS) architecture [1] and components as a Presence server and Xml Document Server (XDMS).

### 1.1 Basic of IMS architecture

The goal of the IMS (Ip Multimedia Subsystem) [1] architecture is to define a model that separates the services offered by fixed-line, mobile, and converged service providers from the access networks used to receive those services. The IMS architecture is based on SIP protocol to enable session establishment between at least two IMS clients or between IMS clients and services. IMS architecture provides some advantages like single sign-on, user subscription management, session management, routing, service trigger, interaction with existing NGN service enablers, roaming, security, bearer control and unified charging and billing.

### 1.2 IMS Presence Server

The Role of Presence Server is to make available presence information of different systems through standard interfaces, in an optimized way. The Presence Server enriches network services or applications by providing dynamic presence information and group management capabilities.

In IMS network, the Presence Server complies with the OMA Presence Data Model [2] and supports the following:

- Standard interfaces to access Presence Server from the IMS Core Network
- Notification on presence change
- Privacy management
- Resource list management
- Optimization of network use

### 1.3 Xml Document Server (XDMS)

The main goal of the XDM Server [3] is to enable authorized end-users to access and handle data such as contact lists and profiles which are used by different applications, such as Presence, Instant Messaging... The data is stored as XML documents in repositories managed by the XDMS.

The XDMS complies with the OMA XML Document Management (XDM) specification [3] and supports the following:
- Storage of contact lists in a common document repository,
- Management functionalities to add, edit, delete contact lists, search, …

- Notification of client upon stored data changes (User Agent Profile management).

1.4 Paper organization

In the section 2 is described the related works on sensor infrastructures. On section 3 we describe our solution "Sensor Search Engine Enabler" to index sensors by type and by location in IMS network to build mash-ups convergent applications using sensors data. A use case study about sensor widgets displayed on a television connected to Internet is presented in the section 4. The conclusion and future works are described in the section 5.

## 2. Related work

On the web, Kansal et al. [4] propose a SenseWeb infrastructure. This infrastructure provides a database responsible on indexing the sensor characteristics and other shared resources in the system to enable applications to discover what is available for their use. The access to the database is based on proprietary web Service API. They didn't provide an open interface for the IMS environment. This solution didn't also notify users or applications when a sensor is not available to be used. Krishnamurthy [5] and Krishnamurthy et al. [6] proposes TinySIP gateway to access wireless multimedia sensor based service. TinySIP gateway is based on SIP (Session Initiation Protocol) which is a standard used in IMS architecture. This gateway maps the SIP messages and MicaZ sensor nodes. TinySIP supports interactions with wireless sensor nodes using publish/subscribe, instant messaging and session semantics. Tomic and Todorova [7] investigate how SIP and ZigBee sensors could interwork in heterogeneous networking scenarios, unifying publish/subscribe/notify concepts of SIP presence service and the discovery and binding concepts ZigBee sensors. To study how the information is conveyed from Wireless Multimedia Sensors to IMS presence server, El Barachi et al. [8] propose a presence based architecture for integration. It is an extension of 3GPP standard presence architecture. Rhee et al. [9] have studied a mobile service which is based on sensors and SIP-specific event notification mechanism. This approach provide a mobile service based on dynamic user's location and made a simple testbed for mobile streaming service using RFID sensors and emulated SIP-based IMS. ARAKI et al. [10] describe a connecting sensor networks to IMS in order to provide new services and a prototype implementation of the proposed system for collecting power consumption data on each devices. Open Geospatial Consortium [11] has specifying interfaces and metadata encodings in order to enable to integrate sensor webs into the information infrastructure in real time.

In all these contributions, authors have solved either availability or identification or the location of the sensors but not all three together. Moreover, the security aspects when sensors are exposed on the Internet have not been studied by authors.

On the present work, we propose a search engine enabler which can be integrated as IMS Application Server on IMS network. The results of the sensor indexation of the sensor search engine are used by Web and/or IMS applications to build convergent mash-ups and multimedia applications.

## 3. The proposed approach

In this section, we present a solution: IMS based Sensor Search Engine Enabler (iSSEE). iSSEE is in charge to provide applications or phone IMS clients, lists of sensors connected to IMS core network in order to collect data (audio, image, video, temperature, …). These lists are built by classifying sensors by types and by locations. Using data provided both by sensors and the web, use cases of mash-ups convergent applications are described.

3.1 Sensor Search Engine Enabler

Independent of the access mode, when a sensor IMS based terminal is connected to internet, it performs a REGISTER request to IMS core network in order to be located as any IMS or SIP phone. The request should contain a public identity to be registered and a home domain name S-CSCF (Serving Call Session Control Function ) [1] which contains the registrar database.

3.1.1 Context

When it is registered in the network and has already published its Presence in the presence server, the sensor is considered as a device phone because the S-CSCF has no information which makes difference between the two types of clients. So, it is difficult for an application or IMS client to use the data provided by these sensors since it doesn't know if it is a sensor or a phone, the type of sensor and its geographical location.

3.1.2 iSSEE solution

To provide IMS applications (Figure 1) or IMS mobile devices, a list of available sensors with information about types and locations, it is necessary to add some information in the sensors SIP requests. We propose to add new headers to identify the type of sensor (sensor-type) and its location (latitude and longitude). After successful registration, the S-CSCF will check the downloaded filter criteria of the registered user (in our case, the sensor name is sensorA). The iSSEE needs to know that the sensor is now been registered and is therefore available. To inform the iSSEE, filter criteria have been configured to trigger all the REGISTER requests that originate from sensorA's public sensor identity and containing a header "sensor-type".

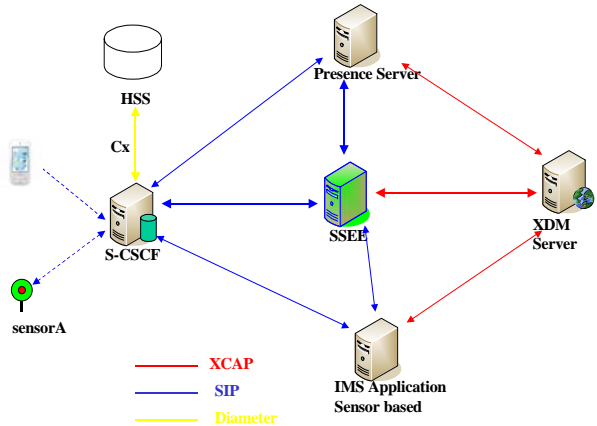

Figure 1 – IMS Sensor Search Engine Enabler

To expose their ability to communicate with other entities, sensor IMS clients send SIP PUBLISH requests to the presence server.

When the IMS client sends a REGISTER request, the sensor-type values (temperature, pressure, camera, humidity, audio) and location information values will be collected and stored by iSSEE in XDMS server.

Since sensors presentities are published in the presence server, the Sensor Search Engine Enabler watchers, subscribe to all sensors, builds sensors lists classified by sensor types and with their locations. The lists are stored in XDM Server. Any IMS application which needs to provide services using sensors can subscribe to these lists of sensors in order to be informed, dynamically, about the connected sensors types and locations. Due to the filter criteria (Figure 2), the S-CSCF will generate a third-party REGISTER request and send it to iSSEE whenever the user performs a successful registration (Figure 3).

```
<InitialFilterCriteria>
<Priority>1</Priority>
<TriggerPoint>
 <SPT>
   <ConditionNegated>0</ConditionNegated>
   <Group>1</Group>
   <SIPHeader>
       <Header>Sensor-type</Header>
       <Content>*</Content>
   </SIPHeader>
 </SPT>
 <SPT>
   <ConditionNegated>0</ConditionNegated>
   <Group>1</Group>
   <SessionCase>0</SessionCase>
 </SPT>
</TriggerPoint>
<ApplicationServer>

<ServerName>sip:issee@192.168.130.76:5050</ServerName>
      <DefaultHandling>0</DefaultHandling>
 </ApplicationServer>
</InitialFilterCriteria>
```

Figure 2 – Initial Filter Criteria

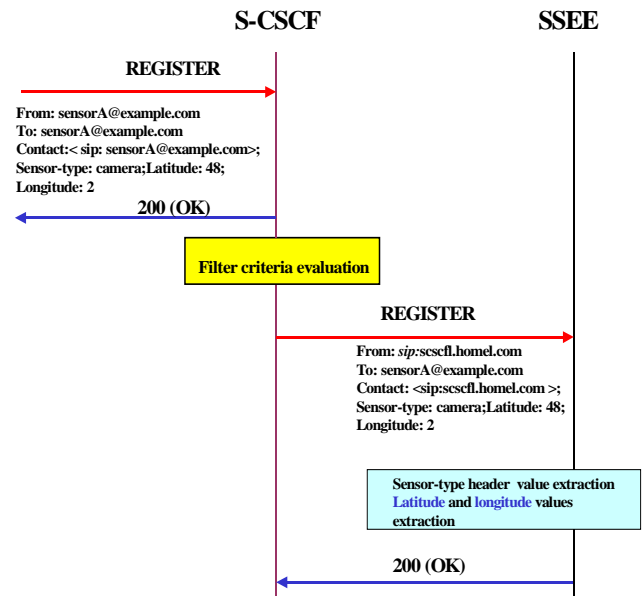

Figure 3 – Third part registration by S-CSCF to iSSEE

This REGISTER request is destined to the iSSEE at issee.home1.com, as indicated in the request URL (Figure 4). The To header includes the public user identity of sensorA, as this is the URI that was registered. The S-CSCF indicates its own address in the From header, as it is registering sensorA's public user identity on behalf of sensorA. Furthermore, the S-CSCF indicates its own address within the Contact header:

> *REGISTER sip:issee.home1.com SIP/2 . 0*
> *Via: SIP/2 . 0/UDP scscfl.home1.com;branch= z9hG4bK-0--1-8321219513718260*
> *Max-Forwards: 70*
> *From:        <sip:scscfl.home1.com>;tag=2-0--1-963621852*
> *To: <sip:sensorA@home1.com>*
> *Contact:     <sip:scscfl.home1.com>;expires=3600; Sensor-type: temperature; Latitude: 48;Longitude: 2*

*Call-ID: 03775315*
*CSeq: 70 REGISTER*
*Content-Length: 0*

Figure 4 – Register request from S-CSCF to iSSEE

This ensures that the iSSEE never routes directly to sensorA's UA (User Agent), but will always contact the S-CSCF first. Before sending back a 200 (OK) response for this REGISTER request to the S-CSCF, the iSSEE collects the "sensor-type" header contents and "longitude" and "latitude" values. If iSSEE needs more information about sensorA's registration state, it can subscribe to the registration-state information of sensorA.

The iSSEE watchers SUBSCRIBE to all presentites containing a "sensor-type" header parameter published on IMS presence server. For each sensor, the iSSEE creates an XML document, identified by sip address and containing information about sensor type, longitude, latitude name of the street address of the location (using latitude, longitude and maps) in addition to standard information. This document is stored in XDM Server database and removed if the sensor presence/register is not refreshed. The iSSEE create groups of sensor documents classified:

- by location – by country or by town or streets, etc.
- by type of sensor - temperature, pressure, camera, etc.
- by pertinence - near known monuments, zoo, Niagara fall, etc.
- by news - based on news events.

All these groups of documents are accessible by IMS applications which subscribe to documents changes in order to provide new services.

The figure 5 shows different functions to search connected sensors to IMS core network:

1 - Adding sensor-type, latitude and longitude header parameters in "contact" header field;

2 - sensor-type trigger point to iSSEE – a third part register is initiated when a sensor-type header parameter is found in the REGISTER request;

3 - Search engine - It orchestrates the search, subscribe and indexing functions and stores documents on XDM server. When the header parameter "sensor-type" is detected on REGISTER request by the S-CSCF, a third part REGISTER is performed to iSSEE. The Search engine subscribes to the sensor UA (User Agent) changes, extracts sensor-type, latitude and longitude headers values, subscribes to published sensor, create a sensor xml document on XDMS database, add the sensor in a group of sensors or create a new group, and subscribes to sensors groups;

4 - Subscribe to S-CSCF – subscribe to the registration-state information of sensors;

5 - sensor-type, latitude and longitude headers parameters values extraction – extracts the headers parameters values of sensor-type, latitude and longitude;

6 - Subscribe to published sensor – Subscribe to sip sensor URI on the presence server;

7 - Create sensor xml document – create an xml document for each sensor URI;

8 - Create group xml document – Create document by location, type, pertinence and actuality events;

9 - Subscribe to sensors group – function which subscribes to group documents;

10 – Store document on XDM server – all documents are stored on XDM server to be used by IMS services.

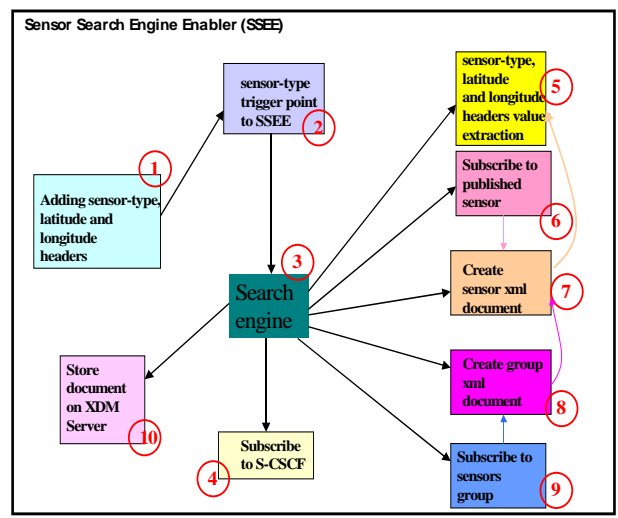

Figure 5 : IMS Sensor Search Engine Enabler (iSSEE)

### 3.1.3 Advantages

This solution allows dynamically indexing by type and by location, all sensors connected to IMS network. Using XML documents stored in XDM Server, which describe the sensors, service providers or Telco have opportunities to provide novel multimedia applications to end users.

### 3.2 Mash-up and convergent application

In this section, we show how we can use iSSEE to built IMS mash-up convergent applications based on distributed sensors.

### 3.2.1 Definitions

An IMS convergent application is an application which uses more than one protocol interfaces. An example of popular application is "click to call". It uses protocols HTTP and SIP. The application receives an HTTP request and initiates a SIP session with the callee phone. When the SIP session is established, an HTTP response will be sent to the HTTP browser. An other example is IMS core operating : when a SIP REGISTER request is received by the S-CSCF, it performs a Diameter Sh request to access some data on the HSS. A good example on convergent applications is those based on JSR289 [12] which can receive both HTTP and SIP requests in the context of same application session.

IMS Mash-ups involve something more than just IMS convergent applications. An IMS mash-up describes an application that combines content or service applications from several heterogeneous data. For example, from analysis of news gathered by a Web search engine, an IMS application enriches information by adding contents (audio, video, image, …) collected by sensors in connection with the information.

3.2.2 IMS Mash-up convergent application

An IMS mash-up convergent application is an application built using at least two protocols and combines contents from several heterogeneous data. Figures 6 and 7 show an example of IMS mash-up convergent application. On Figure 6, the sensor based IMS application has an interface with web servers to become convergent. It shows how we use heterogeneous data on an IMS mash-up convergent application based on textual information enriched by multimedia content. The IMS mash-up subscribe first to textual information in the web (google for example). When it receives a notification from the web, the mash-up application looks for sensors which match with the textual information. When a sensor is found, the mash-up convergent application establish a session (Figure 7) with the sensor by sending INVITE requests. After establishing a SIP session, data are transferred from sensors to be stored in multimedia databases. These data and other information collected from the Web are combined to provide rich information corresponding to the end-user profile.

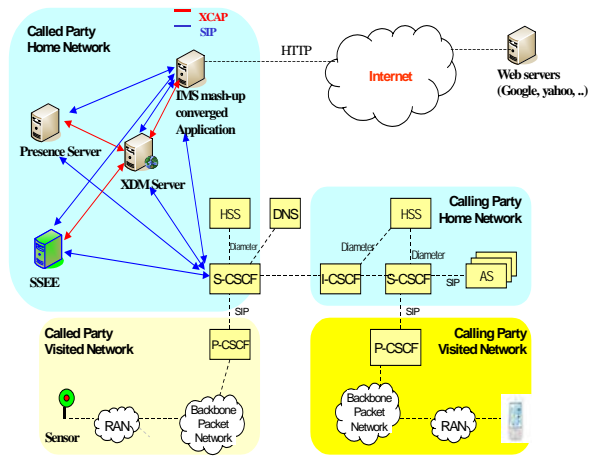

Figure 6 - IMS mash-up convergent application

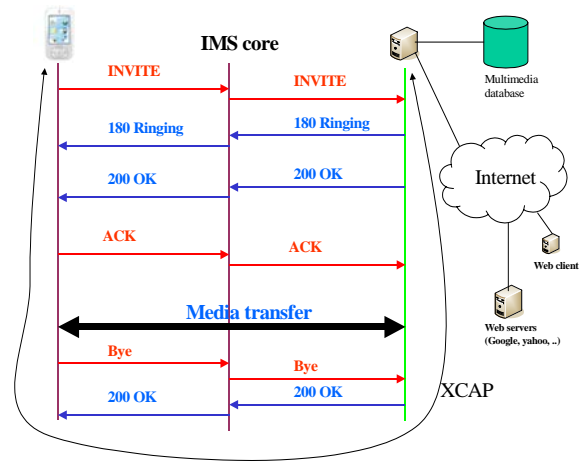

Figure 7 – IMS application data store

- **IMS client**

An IMS or Web end user can also access directly to data provided by lists of sensors built by iSSEE (Figure 9). This is done by requesting a sensor public SIP URI in the same way then a SIP phone.

## 4. A user case study

After the discussion of our proposal, we will introduce some use cases in Hypermedia domain. Hypermedia is a logical extension of the term hypertext in which graphics, audio, video, plain text and hyperlinks intertwine to create a generally non-linear medium of information.

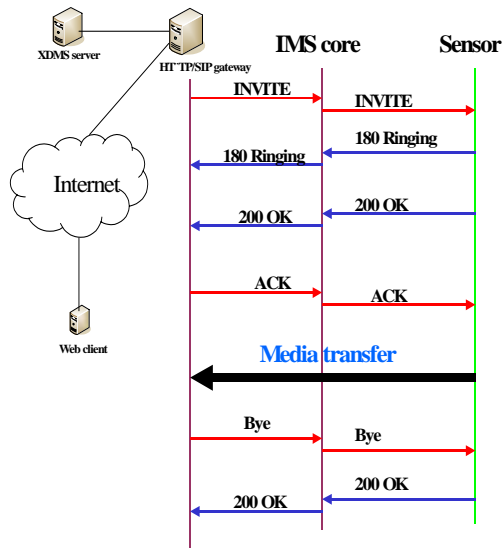

Figure 8 – Http client requesting a sensor

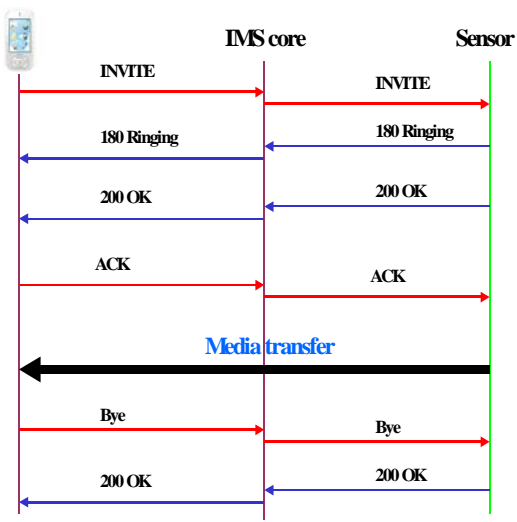

Figure 9 – SIP phone requesting a sensor

End users need more links, more interactivity, more easiness in exploring massive media and associated knowledge, and also more involvement on the creation and production side. Distributed sensors and internet of things provide an opportunity to end users to create/enrich contents and applications for themselves or shared with communities.

The study case we will describe is about sensor widgets displayed on a television connected to Internet. Different application can be provided to end user using sensor widgets:

- Display sensor widgets by clicking on the remote control button and select the list of sensors by type, location, etc.
- Automatic displaying, in real-time, sensor widgets related to the program viewed by the end user. The number of displayed widgets is configured by the user;
- Display actuators widgets by clicking on the remote button.

### 4.1 Displaying lists of sensor widgets

By clicking on the remote control menu, lists of sensors are displayed on the TV screen and classified by location (Figure 10) and/or by type. A user can select one or several sensor widgets to be displayed permanently or at defined time.

### 4.2 Automatic real-time displaying widgets

If rich meta-data are synchronized with broadcast TV flow, sensor widgets which match with sensor meta-data can be displayed automatically on the TV (Figure 11) to enrich the TV program by adding heterogeneous information. The IMS TV client establish sessions with each sensor to display in real-time its data in widgets.

### 4.3 Displaying actuator widgets

Some sensors can contain actuators mechanism which allow user from his television to take control of the camera sensor in order to change its position and perform zooming. In this case and in the same SIP established session, a user manipulates the actuator to see the media transferred by the sensor.

## 5. Conclusion and Future works

In this paper is presented an architecture of a sensor search engine enabler (iSSEE). The iSSEE allows to create lists of available sensors stored in XDMS database. It creates XML documents classified by type of sensors and by location. Any mash-up convergent applications, subscribing to the lists can use the data provided by these sensors to enrich applications. A phone IMS based can also establish a SIP session to any sensor found in the lists in order to get data.

As a future work, a sensor search engine enabler will be implemented as it is described above. An IPTV interactive service using widget sensors will be also implemented and using a TV HTML browser installed on set-top box or on TV.

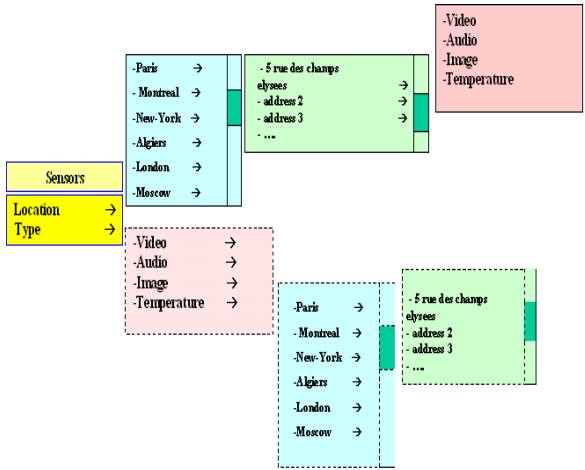

Figure 10 – Displaying lists of sensor widgets

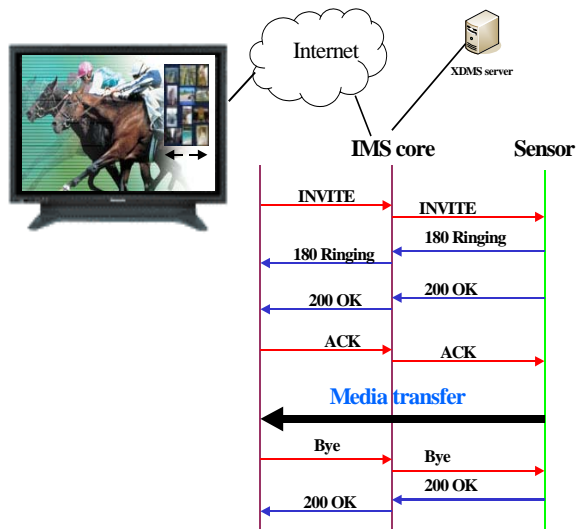

Figure 11 – Automatic displaying widgets

**Abdelkader Outtagarts** is a research engineer in Alcatel-Lucent Bell Labs France's Applications domain. As a member of the Hypermedia department, his research interests are focused on applications impacting end user in collaborative context, user knowledge extraction and multi-agents systems. Abdelkader has worked in Alcatel-Lucent CTO application as software engineer and security engineering manager of IMS products (Application Server, Presence Server, XDM server). Abdelkader receives his Master in heat transfer in 1990 and his Ph.D in Thermal and Energy in 1994 from the National Institute of Applied Science of Lyon (France). He also received a Master in software engineering in 1999 from High School Technology at Montreal (Canada). He has 12 years experiences on Internet technology and applications.

**Olivier Martinot** leads the HyperMedia department in the Applications Domain team at Alcatel- Lucent Bell Labs in Villarceaux, France. His primary research interests include advanced applications enabling end-users' easy access to media and information, based on hypermedia links created using metadata, personalization, and context. He is an aeronautic and space engineer graduated from the French Sup'Aéro Engineering School in Toulouse, France.